\documentclass[fleqn,twoside]{article}
\usepackage{espcrc2}
\usepackage{graphicx}
\usepackage{ams}

\begin{document}
\title{Wroc\l aw neutrino event generator
}
\author{Jaros{\l}aw A. Nowak \address{Institute of Theoretical Physics, Wroc\l aw University,\\ Pl. M. Borna 9, 50-204 Wroc\l aw, Poland}}
\runtitle{Wroc\l aw neutrino event generator} \runauthor{J. A. Nowak}

\begin{abstract}
A neutrino event generator developed by the Wroc\l aw Neutrino Group is described. The physical
models included in the generator are discussed and illustrated with the results of simulations. The
considered processes are quasi-elastic scattering and pion production modelled by combining the
$\Delta$ resonance excitation and deep inelastic scattering.
\end{abstract}

\maketitle

\section{Introduction}

Many of the neutrino experiments now operating or in preparation are those with artificial neutrino
beams. The beam energy ranges from $0.5$ to about $30~GeV$ (MINOS, CNGS, T2K). The uniform
theoretical description of neutrino interaction within this region is problematic, because three
different dynamics: quasielastic(QE), resonance excitation(RES), and deep inelastic scattering(DIS)
 must be taken into account. Especially in the region of about
 $1~GeV$ they contribute with comparable strength and the problem
of the overlap between the DIS formalism and the resonances
production model appears.

Up to now most of the Monte Carlo codes use the Rein-Sehgal model of the resonances production.
However, another approach has been proposed~\cite{NJS05} which makes use of the postulated
quark-hadron duality in neutrino reactions. One can assume that the contributions from higher
resonances are averaged by the deep inelastic scattering structure functions and only the dominant
$\Delta$ resonance has to be treated separately. In this approach the exclusive inelastic final
states must be obtained from the DIS formalism even in the resonance region $W \in
(M+m_\pi,~2~GeV)$, where the most important contribution comes from pion production channels.

\section{Generator description}

Dynamical models included in the generator are described below. For the quasielastic scattering the
usual Llewellyn Smith~\cite{LS} model is used with BBBA05 form factors~\cite{BBBA05}. For elastic
scattering new strange form factors are taken into account~\cite{ABC02}. For antineutrino CC
interaction there are three additional channels with hyperons $\Lambda,\ \Sigma^+,\ \Sigma^0$
production($|\Delta Y|=1$). The $\Delta$ resonance is described by the form factors as
in~\cite{ASV98}, and for DIS we use parton distribution functions GRV94~\cite{GRV95} with
modification proposed by~\cite{BY02}. In order to get the final state from the inclusive DIS cross
section we use model of fragmentation based on the LUND model, implemented in the PYTHIA6
generator. It is assumed that the interaction occurs on a separate constituent of the nucleon. The
fragmentation procedure was described in~\cite{JNS06}. In order to get good agreement with
experimentally measured charged particle multiplicity~\cite{Z83} we have changed a few PYTHIA6
parameters. The final values, fine-tuned for specific procedure of choosing the interacting parton
inside the nucleon, are : PARJ(32)=$0.1~GeV$, PARJ(33) = $0.5~GeV$, PARJ(34) = $1~GeV$, PARJ(36)=
$0.3~GeV$, MSTJ(17) = 3. The definitions and default values of the parameters one can find in
PYTHIA manual~\cite{SJO1}.

We use a procedure which gives exclusive cross sections for each single pion production channel
(SPP):
\[
\frac{d\sigma^{SPP}}{dW}=\frac{d\sigma^{\Delta}}{dW}\left(1-\alpha(W)\right) +
\frac{d\sigma^{DIS}}{dW}F^{SPP}(W)\alpha(W)
\]
where $F^{SPP}$  is the percentage of the given SPP channel within the overall DIS cross section.
The function $\alpha(W)$ defines transition between
\begin{figure}
  \includegraphics[scale=0.5]{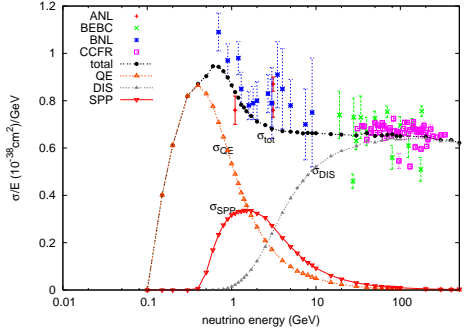}\\
  \vspace*{-1cm}
  \caption{Cross section for neutrino scattering $\nu N \to \mu^- X$.}\label{cr_nu}
\end{figure}
\begin{figure}
  \includegraphics[scale=0.5]{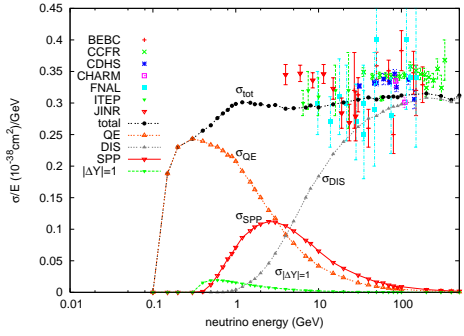}\\
  \vspace*{-1cm}
  \caption{Cross section for antineutrino scattering $\bar \nu N \to \mu^+ X$. $|\Delta Y|=1$ denotes the sum
  of cross sections for quasielastic hyperons production.}\label{cr_anu}
\end{figure}
\begin{figure}
  \includegraphics[scale=0.5]{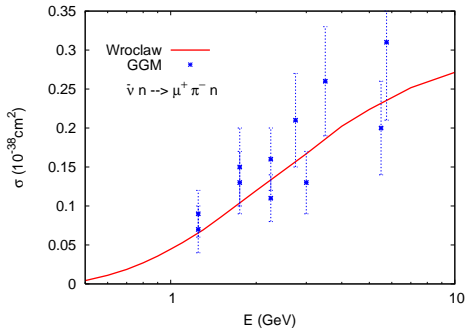}\\
  \vspace*{-1cm}
  \caption{Cross section for $\bar \nu n \to \mu^+ \pi^- n$.}\label{cr_anu_spp}
\end{figure}
\begin{figure}
  \includegraphics[scale=0.5]{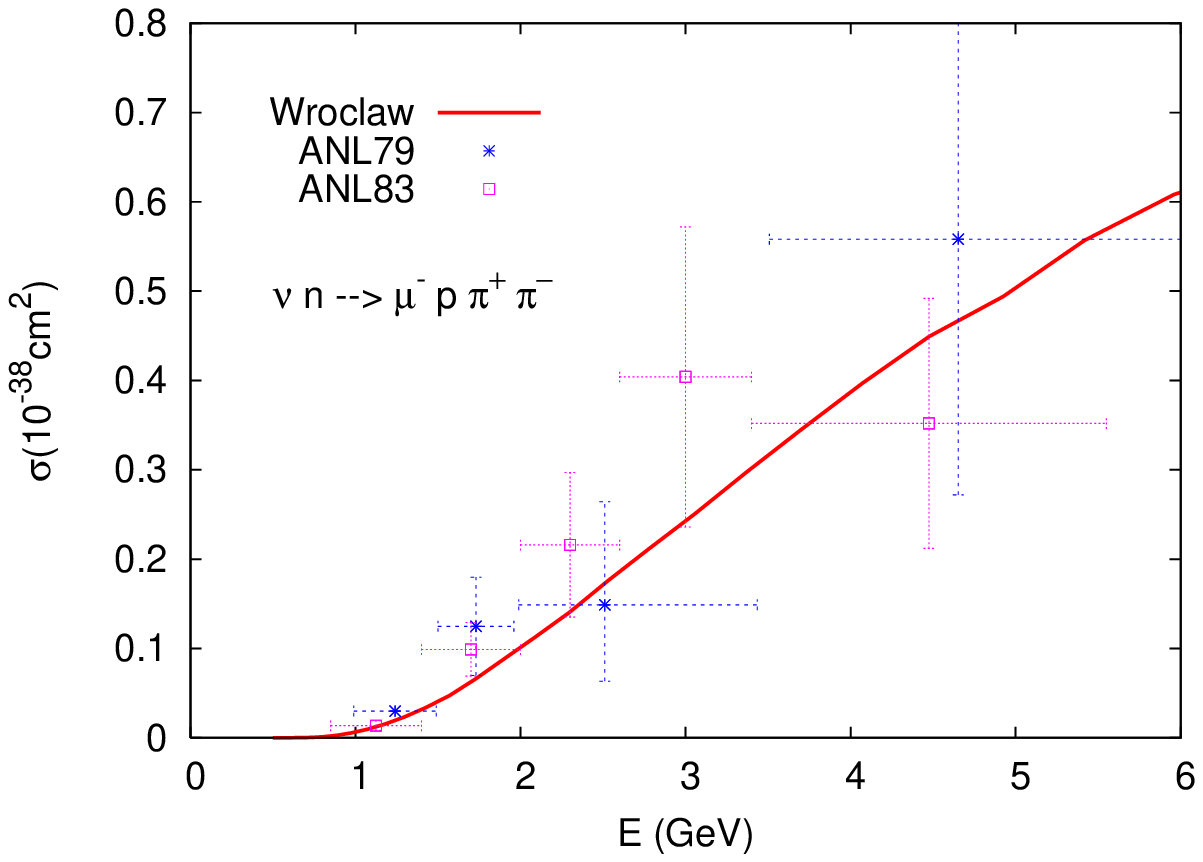}\\
  \vspace*{-1cm}
  \caption{Cross section for double-pion production: $\nu n \to \mu^- p  \pi^+ \pi^-$.}\label{cr_2pp}
\end{figure}
\begin{figure}
  \includegraphics[scale=0.5]{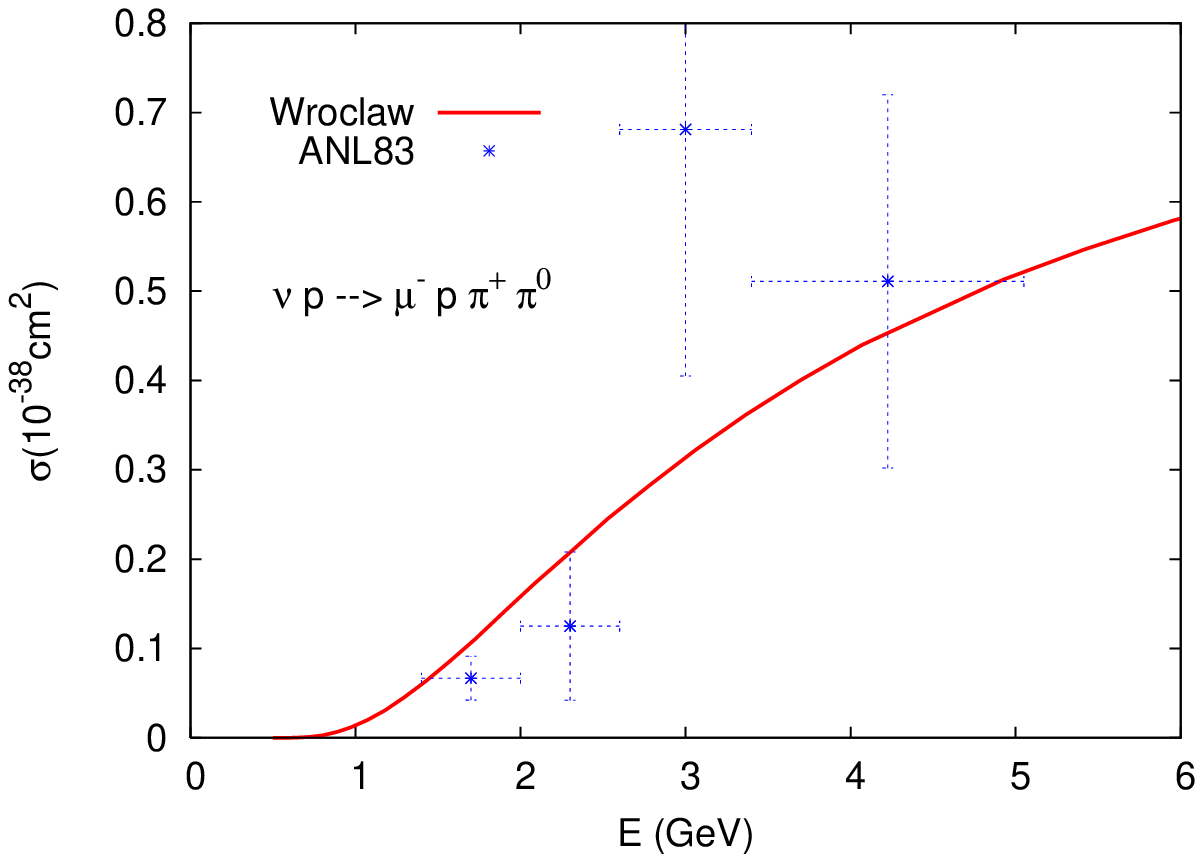}\\
  \vspace*{-1cm}
  \caption{Cross section for double-pion production: $\nu p \to \mu^- p  \pi^+ \pi^0$.}\label{cr_2pp2}
\end{figure}
\begin{figure}
  \includegraphics[scale=0.5]{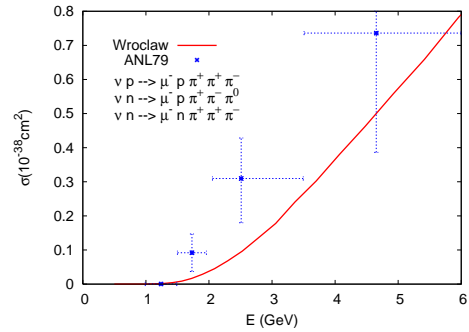}\\
  \vspace*{-1cm}
  \caption{The sum of the three triple-pion production cross sections.}\label{cr_3pp}
\end{figure}
two SPP models: mediated by the $\Delta$ resonance and the inelastic one, and accounts
for the non-resonant background.

\section{Results}
Many of the results from the Wroclaw Neutrino Generator have already been presented~\cite{JNS06}.
The basic results for CC neutrino and antineutrino scattering off nucleon are shown in
figs.~\ref{cr_nu},~\ref{cr_anu}). The cross section for SPP channel induced by antineutrino is
shown in fig.~\ref{cr_anu_spp}.  In figs.~\ref{cr_nu},~\ref{cr_anu}, and~\ref{cr_anu_spp} only SPP
events with $W<2~GeV$ are taken into account. The cross section for two channels with two final
pions, for which experimental data exists are presented in figs.~\ref{cr_2pp}, \ref{cr_2pp2}, and
the sum of the cross sections for three pions production is given in fig.~\ref{cr_3pp}. The
simulations results agree with the experimental data for single and double pion production, but
some discrepancy occurs for triple-pion production.


\section*{Acknowledgements}
The author would like to thank J Sobczyk, C Juszczak, and K Graczyk for many fruitful conversation
and for participating in this project at various stages. The author was supported by KBN grant
105/E-344/SPB/ICARUS/P-03/DZ211/2003-2005.

\end{document}